\documentclass[aps,prc,showpacs,showkeys]{revtex4}

\usepackage{graphicx,amsmath,amssymb,bm}

\newcommand{\eq}[1]{Eq.~(\ref{#1})}
\newcommand{\be}{\begin{equation}}
\newcommand{\ee}{\end{equation}}
\newcommand{\bea}{\begin{eqnarray}}
\newcommand{\eea}{\end{eqnarray}}
\newcommand{\bfr}{\bf r}

\def\bfq {{\bf q}}\def\bfb{{\bf b}}\def\bfB{{\bf B}}
\def\bfr {{\bf r}}\def\bfR {{\bf R}}\def\bfP{{\bf P}}
	
\newcommand{\bfkappa}{\mbox{\boldmath $\kappa$}}

\def\bfk{{\bf k}}
\def\bfp{{\bf p}}  

\def\bfz{{\bf z}}

\begin{document} 


\title{\hskip 9.5cm NT@UW-09-18\\ Electromagnetic Form Factors and
  Charge Densities From Hadrons to Nuclei} 
\preprint{NT@UW-09-??}
\author{ 
Gerald   A. Miller}

\affiliation{Department of Physics,
University of Washington\\
Seattle, Washington 98195-1560
}

\begin{abstract} A simple exact covariant   model
 in which  a scalar particle $\Psi$ 
is  modeled as a bound state of
two different  particles  is used to elucidate
relativistic aspects of electromagnetic form factors $F(Q^2)$. 
The model  form factor is computed  using an exact  covariant calculation of
the lowest-order triangle diagram. The
light-front technique 
of integrating over the minus-component of the virtual momentum gives
the same result and is the same as  the one obtained originally by 
Gunion {\textit et al.}   by using time-ordered perturbation theory in
 the infinite-momentum-frame.  The 
 meaning of  the transverse density $\rho(b)$ is explained by providing a  general
 derivation, using three spatial-coordinates, of its  relationship
 with the 
form factor. This allows us to
 identify a mean-square transverse size $\langle b^2\rangle=\int
 d^2b\;b^2\rho(b) =-4{dF\over dQ^2}(Q^2=0)$. The quantity  $\langle
b^2\rangle$ is a true measure of hadronic size because of its direct
relationship with the transverse density. We 
 show that the rest-frame charge
distribution is generally not observable by studying the 
explicit failure to uphold current conservation.  
  Neutral systems of two charged constituents are shown to 
obey the conventional lore that the heavier one is generally closer to
the transverse origin than the lighter one. It is argued that the
negative central charge density of the neutron arises, in pion-cloud
models,
 from pions of high longitudinal momentum that 
reside at the center.
The non-relativistic limit is defined
precisely, and    the ratio of the binding energy
$B$ to the mass ${\cal M} $ of the lightest constituent
 is shown to govern the influence of
relativistic effects. We show that 
the exact relativistic formula for $F(Q^2)$ is  the same as 
the
familiar one of the three-dimensional Fourier transform of a square of
a wave function  for very small values of $B/{\cal M}$, but this only
occurs values of  $B/{\cal M}$  
less than about 0.001. 
For masses that mimic the quark-di-quark
model of the nucleon we find that there are substantial relativistic
corrections to the form factor for any value of $Q^2$. A schematic
model of the lowest $s$-states of nuclei is developed.
 Relativistic
effects are found to decrease the form factor for light nuclei but to 
increase the form factor for heavy nuclei. Furthermore, these
lowest $s$-states are likely to be strongly influenced by relativistic
effects that
are order 15-20\%.  
\end{abstract}
\maketitle

\section{Introduction}

The  text-book  interpretation of nucleon electromagnetic  form factors is  that their three-dimensional
 Fourier transforms are measurements of the charge and magnetization
 densities. 
This interpretation is deeply buried in our 
 thinking and continues to guide intuition as it has since the days
 of the Nobel prize-winning work of Hofstadter\cite{Hofstadter:1956qs}.
 Nevertheless, the relativistic motion of the constituents of the
 system  
causes the text-book interpretation to be incorrect. 
  
The preceding statement leads to a number of questions, the first
 being: Is the statement correct? If correct, how relativistic does
 the motion of the constituents have to be? Why is it that the relativistic
 motion of the constituents and not that of the entire system that 
causes the non-relativistic approach to fail?  It is probably true that the answers to these questions are displayed within the existing literature. However, 
 obtaining  general clear answers has been sufficiently difficult that posing even the first question of this paragraph would not lead to a  unanimous  answer by all professionals in the field.
 
This paper offers the strategy of using a simple model, a 
generalization of the $\phi^3$ model used by Weinberg \cite{Weinberg:1966jm} 
to illustrate advantages of using 
the infinite momentum by choosing frame that was
used 
 by Gunion {\it et al.}  \cite{Gunion:1973ex}   to explore form factors and hadronic interactions at 
 high-momentum transfer. We take the 
interaction Lagrangian density to be of the form $g \Psi\phi\xi$ in
which all of the fields are bosons.The $\Psi$ particle of mass $M$
represents the bound state of the two different constituents
$\phi,\xi$ of masses $m_1,m_2$ respectively. Thus the $\Psi$
represents
 the hadron or nucleus with the $\phi,\xi$ representing the quark or nucleonic constituents. Mass renormalization effects are ignored.  
We can choose either or both of the constituents to be charged and thus discuss charged and neutral $\Psi$ particles.

The motivation to pose questions regarding the meaning of
electromagnetic form  factors  at this point in time arises from recent
experimental work, especially the discovery that the ratio of the
proton's electric to magnetic Sachs form factors $G_E/G_M$ drops
rapidly (please see the reviews \cite{reviews}) 
and from our recent finding \cite{Miller:2007uy}, based on
measurements and the
 use of the transverse density  that the charge density at the
 neutron's center is negative. The nucleon transverse density 
$\rho(b)$, the two-dimensional Fourier transform of $F_1$ is the
 infinite momentum frame charge density\cite{notation}     
    located at a transverse separation $b$ from the center of
    transverse
    momentum\cite{soper1,mbimpact,diehl2,Carlson:2008zc}. 
This quantity has a direct relationship to matrix element of a 
density operator. The usual three-dimensional Fourier transforms of $G_E$ and $G_M$ do not because 
the initial- and final-state nucleons have different momentum, and therefore different wave functions.
This is because the  
relativistic boost operator that  transforms a nucleon at  rest  into  a moving one changes
the wave function in a manner that depends on the momentum of the
nucleon.  However, we expect that there are non-relativistic
conditions for which the text-book interpretation is correct. 
We aim
 to explore those conditions
 by choosing appropriate values of the masses $m_1,m_2,M$.

Here is the outline we follow.  The form factor for the situation in which the $\Psi$ and $\phi$ carry a single unit of charge, but the
$\xi$ is neutral, is computed using an exact  covariant calculation of
the 
lowest-order triangle diagram in Sect.~II. This is followed by a  another derivation using the  light-front technique of integrating over the minus-component of the virtual momentum in Sect.~III that obtains the same form factor. This is also the result obtained originally by 
 \cite{Gunion:1973ex}   by using time-ordered perturbation theory in
 the infinite-momentum-frame IMF. Thus three different approaches
 yield the same exact result for this model problem. Any approximation
 that does not yield the same form factor is simply not
 correct. The asymptotic limit of
 very high momentum transfer $Q^2$ is also studied. The next section (IV) explains 
 the transverse density of the model, its central value, a general
 derivation of its relationship with the form factor using three
 dimensional spatial coordinates and the meaning of hadronic radii.
Section~V  displays the spatial wave function in terms of three
spatial coordinates. Section~VI shows that the rest-frame charge
distribution is generally not observable. Section~VII is concerned
with the question of whether neutral systems of two constituents
obey the conventional lore that the heavier one is generally closer to
the origin than the lighter one. The non-relativistic limit is defined
and applied to a variety of examples in
Section~VIII. The exact formula for the form factor morphs into the
familiar one of the three-dimensional Fourier transform for
sufficiently large values of the constituent mass divided by the
binding energy of the system. Examples that are motivated by the pion,
deuterium, nucleon and heavy nuclei are provided. This work is
summarized in Section IX.

\section{Exact form factors using a simple model}
\begin{figure}
\unitlength.9cm
\begin{picture}(14,8.2)(.5,-.8)
\includegraphics{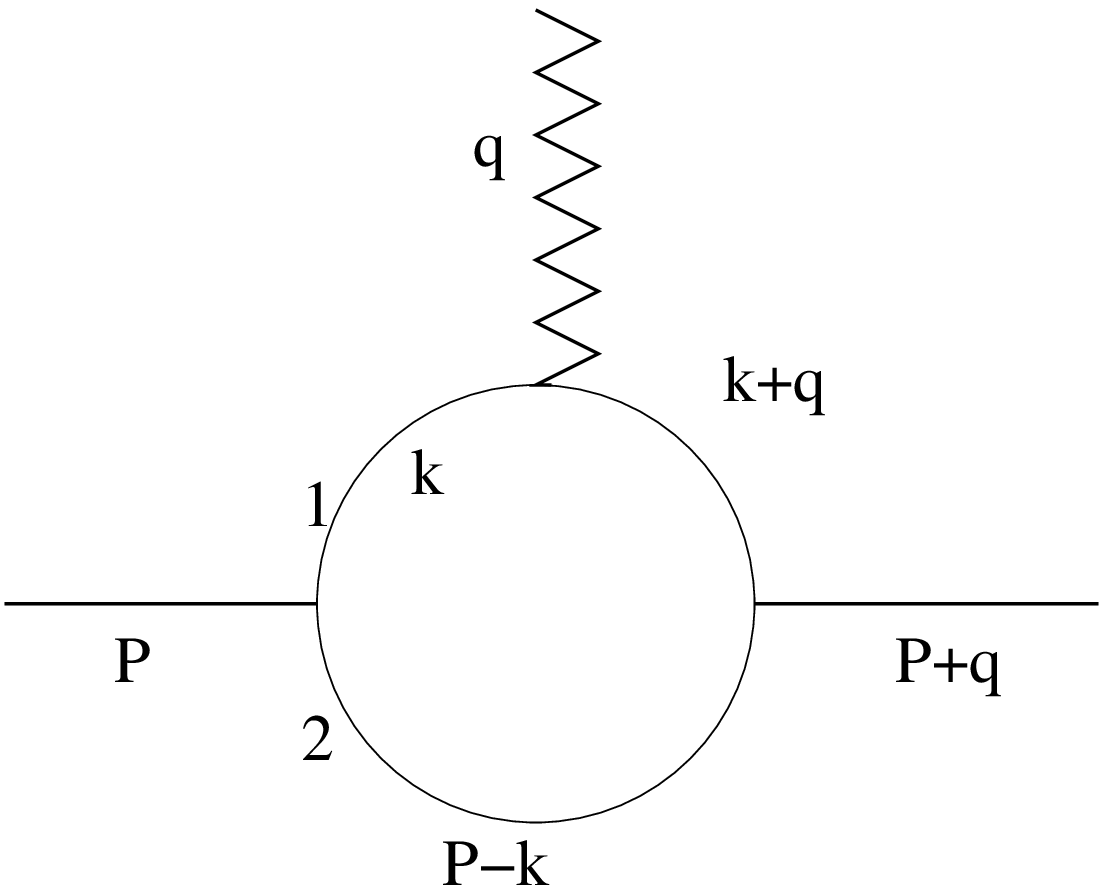}
\end{picture}
\caption{\label{ff} 
Feynman diagram for the form factor with the photon coupling
  to the $\phi$ particle of mass $m_1$. The initial and final hadron
  $\Psi$ carry momentum, $P$ and $P+q$. The $\xi$ is a spectator. }
\end{figure}
The model Lagrangian density is given by  $g\Psi\phi\;\xi$ where $\Psi,\phi$ and
$\xi$ represent three scalar
fields of masses $M,m_1$ and $m_2$ respectively and $g$ is a coupling constant.
 One can take two or three of these fields to carry charge to make up
 a system of definite charge (including the case when the hadron
 $\Psi$ is  neutral). The effects of mass renormalization are not
 considered here because we aim to use a simple model to provide
 easily calculable examples 
and illustrate specific points. The condition  $m_1+m_2>M$ is used to
insure  that the
hadron $\Psi$ is stable.
 
  We start with the situation in which 
 the $\Psi,\phi$ each  carry a single positive charge and
 $\xi$ is neutral.  The form
 factor $F(q^2)$ for a space-like incident photon  of four-momentum
 $q^\mu\; (q^2<0,Q^2=-q^2)$, incident on a target $\Psi$ of
 four-momentum $P^\mu$ 
interacting with the
$\phi$ of mass $m_1$
 is given, to lowest order in $g$, 
 by the single Feynman diagram of Fig.~1. We take the model electromagnetic
 current $J^\mu$ (in units of the proton charge) as given by 
\bea J^\mu=\phi\stackrel{\leftrightarrow}{\partial}^\mu\phi,\eea
and find
\bea \langle P+q|J^\mu(0)|P\rangle\equiv
&&F(Q^2)(2P^\mu+q^\mu)\label{def}\\
&&=-ig^2 \int {d^4k\over (2\pi)^4}
{1\over (k^2-m_1^2+i\epsilon)}(2k^\mu+q^\mu){1\over ((k+q)^2-m_1^2+i\epsilon)}
{1\over ((P-k)^2-m_2^2+i\epsilon)}.\label{feyn1}
\label{diag}\eea
Proceed by combining the denominators using the Feynman procedure and shifting the origin of the convergent  integral   to find
\bea
F(Q^2)(2P^\mu+q^\mu)=-i2g^2 \int {d^4\kappa\over (2\pi)^4}\int_0^1dx \int_0^{1-x}dy{q^\mu(1-2y)+2P^\mu x\over[\kappa^2-{\cal M}^2+i \epsilon]^3},\eea
where
${\cal M}^2=Q^2y(y+x-1)+M^2x(1-x)-m_1^2(1-x)-m_2^2.$ Use
\bea \int {d^4\kappa\over (2\pi)^4}
{1\over (\kappa^2-{\cal M}^2+i\epsilon)^3}=-i{\pi^2\over (2\pi)^4}{1\over2}{1\over {\cal M}^2}
\eea
to find
\bea F(Q^2)(2P^\mu+q^\mu)=-{g^2\over 16\pi^2} \int_0^1dx
\int_0^{1-x}dy{q^\mu(1-2y)+2P^\mu x\over
  Q^2y(y+x-1)+M^2x(1-x)-m_1^2(1-x)-m_2^2x}.\eea
The integral over $y$ can be done in closed form with the result
\bea F(Q^2)(2P^\mu+q^\mu)=4{g^2\over 16\pi^2} (q^\mu+2P^\mu)\int_0^1dx
{ x{\rm Tanh}^{-1}\left[{\sqrt{Q^2}(1-x)\over\sqrt{4x\; m_2^2+4(1-x)m_1^2-x(1-x)M^2+(1-x)^2Q^2}}\right]\over\sqrt{Q^2}\sqrt{4x \;m_2^2+4(1-x)m_1^2-4x(1-x)M^2+(1-x)^2Q^2}}.
 \eea
The above expression shows that current conservation is satisfied and
that the form factor can be obtained from any component of the current
operator $J^\mu$.  
The final result for the form factor is 
\bea F(Q^2)={g^2\over 4\pi^2}\int_0^1dx
{ x{\rm Tanh}^{-1}\left[{\sqrt{Q^2}(1-x)\over\sqrt{4x\; m_2^2+4(1-x)m_1^2-x(1-x)M^2+(1-x)^2Q^2}}\right]\over\sqrt{Q^2}\sqrt{4x\; m_2^2+4(1-x)m_1^2-4x(1-x)M^2+(1-x)^2Q^2}}.
\label{final} \eea
This closed form expression is the key result of this paper. It can be
modified to 
describe a variety of different physical situations. 
\section{Infinite momentum frame/Light Front Representation}
We derive the light front representation by starting with the form
factor of \eq{feyn1} and integrating over $k^-$. This procedure is
simplified by choosing $q^+=0$, so that $Q^2=\bfq^2$ \cite{notation}
 and evaluating the $+$ component of the
electromagnetic current operator. In the present Section  the
convention is that
$A^\pm=A^0\pm A^3$ for the four-vector $A^\mu$. Then  \eq{feyn1}
becomes
\bea 2P^+ F(Q^2)=-ig^2\int {d^4k\over (2\pi)^4}
\left
[{2k^+\over
    (k^2-m_1^2+i\epsilon)}
{1\over ((k+q)^2-m_1^2+i\epsilon)}
{1\over ((P-k)^2-M^2+i\epsilon)}\right]
\\
=-ig^2\int {d^4k\over (2\pi)^4} {2k^+\over {k^+}^2(P^+-k^+)}  
 {1\over ( k^--{(\bfk^2+m_1^2)\over k^+}+{i\epsilon\over k^+})} 
 {1\over ( k^--{((\bfk+\bfq)^2+m_1^2)\over k^+}+{i\epsilon\over k^+})} 
 {1\over (P^-- k^--{(\bfP-\bfk)^2+m_2^2)\over P^+-k^+}+{i\epsilon\over
     P^+-k^+})}
\eea 
If we integrate over the upper half of the complex $k^-$ plane we find
a non-zero contribution only for the case $0<k^+<P^+$. Carrying out
the integral leads to 
\bea2P^+F(Q^2)={g^2\over (2\pi)^3}\int d^2\bfk \int {dk^+\over
k^+(P^+-k^+)}
{1\over P^--{\bfk^2+m_1^2\over k^+}- {(\bfP-\bfk)^2+m_2^2\over P^+-k^+}}
{1\over P^--{(\bfk+\bfq)^2+m_1^2\over k^+}- {(\bfP-\bfk)^2+m_2^2\over
    P^+-k^+}}
\eea
Next change variables by defining 
\bea x\equiv {k^+\over P^+},\eea
so that 
\bea F(Q^2)={g^2\over 2(2\pi)^3}\int d^2\bfk \int_0^1 {dx\over x(1-x)}
{1\over P^+P^--{\bfk^2+m_1^2\over x}- {(\bfP-\bfk)^2+m_2^2\over 1-x}}
{1\over P^+P^--{(\bfk+\bfq)^2+m_1^2\over x}- {(\bfP-\bfk)^2+m_2^2\over
    1-x}}.\eea
Further define the relative transverse  momentum 
\bea \bfkappa\equiv (1-x)\bfk-x(\bfP-\bfk)=\bfk-x\bfP,\eea
so that the form factor can be re-expressed as
\bea F(Q^2)={g^2\over 2(2\pi)^3}\int d^2\bfkappa 
\int_0^1 {dx\over x(1-x)}
{1\over M^2-{\bfkappa^2+m_1^2\over x}- {\bfkappa^2+m_2^2\over 1-x}}
{1\over M^2-{(\bfkappa+(1-x)\bfq)^2+m_1^2\over x}- {(\bfkappa+(1-x)\bfq)^2+m_2^2\over
    1-x}}.\label{lf}\eea
This is the  expression obtained in Ref.~\cite{Gunion:1973ex},
by using time-order-perturbation  theory in the infinite momentum frame.
Integration over $k^-$
leads to the same result for this example. 

It is useful to re-express the result \eq{lf} in terms of a wave function $\psi$ with
\bea \psi(x,\bfkappa)\equiv g[M^2-{\bfkappa^2+m_1^2\over x}- {\bfkappa^2+m_2^2\over 1-x}]^{-1} \label{wf}
.\eea In that case
\bea F(Q^2)={1\over 2(2\pi)^3}\int d^2\bfkappa 
\int_0^1 {dx\over x(1-x)}\psi^*(x,\bfkappa+(1-x)\bfq)\psi(x,\bfkappa),
\label{2dft}\eea
as found in Ref.~\cite{Gunion:1973ex}.

The integration over $\bfkappa$ is convergent so we
 carry out the integration over $\bfkappa$ by combining the
propagators and shifting the origin. This gives
 \bea F(Q^2)={g^2\pi\over 2(2\pi)^3}  \int_0^1 {dx
}\int_0^1 dz
{x(1-x)\over (1-x)m_1^2+ x m_2^2-x(1-x)M^2+(1-x)^2\bfq^2z(1-z)}.\eea
The integral over $z$ can be done with the result that
\bea F(Q^2)={g^2\over 4\pi^2}\int_0^1dx
{ x{\rm Tanh}^{-1}\left[{\sqrt{Q^2}(1-x)\over\sqrt{4x\; m_2^2+4(1-x)m_1^2-x(1-x)M^2+(1-x)^2Q^2}}\right]\over\sqrt{Q^2}\sqrt{4x\; m_2^2+4(1-x)m_1^2-4x(1-x)M^2+(1-x)^2Q^2}}\label{exactform}
 \eea
This is  the same as our previous exactly computed result, \eq{final}. Thus
 evaluation in the infinite momentum frame or the equivalent (for this
 model) light front technique of integration over $k^-$ yields the
 exact result.
\subsection{Asymptotic Behavior of the Form Factor}

The limit of very high $Q^2$ is of considerable interest. One wants to
see how the quark counting rules emerge from an exact calculation,
even if the model is very simple. To this end we note, that the
integral \eq{exactform}  can be evaluated exactly in the limit that
$m_1=m_2=m$ with $M=0$. Then measuring $Q\equiv \sqrt{Q^2}$ in units of
$m$ ($Q/m\rightarrow Q$) we find
\bea F(Q^2)={g^2\over
  4\pi^2}\frac{\log ^2\left(\frac{1}{2} \left(Q
   \left(\sqrt{Q^2+4}+Q\right)+2\right)\right)+8}{8
   Q^2}-\frac{\sqrt{Q^2+4} \log \left(\frac{1}{2} \left(Q
   \left(\sqrt{Q^2+4}+Q\right)+2\right)\right)}{2 Q^3}
\eea
so that 
\bea \lim_{Q^2\rightarrow\infty}F(Q^2)=
\frac{\frac{1}{2} \log ^2\left(\frac{1}{Q}\right)+\log \left(\frac{1}{Q}\right)+1}{Q^2}+
\frac{\log \left(\frac{1}{Q}\right)-1}{Q^4}+\cdots.\label{lim}
\eea

Thus the leading asymptotic behavior is 
\bea \lim_{Q^2\rightarrow\infty}
F(Q^2)\sim \frac {{\frac{1}{2} \log}^2 Q^2}{Q^2}.\label{simple}\eea
Thus the power-law fall-off expected from the quark-counting     rules
appears, but it is modified by the presence of the logarithms.
This behavior is not associated with taking $M$ to zero because in all
cases we have $m_1+m_2>M$ as required for the  particle to be stable. Thus the
asymptotic behavior ($Q^2\gg m_1^2,m_2^2$) associated with \eq{simple}
is expected to be  universal for this model. Note however, from \eq{lim} that
the approach to this asymptotic form is very slow.

\section{Electromagnetic form factors measure transverse densities and
transverse radii}
The expression \eq{2dft} is noteworthy because the form factor is
expressed as a three-dimensional integration that involves
momentum-space wave
functions evaluated at different initial and final momenta. If the $(1-x)$ factor
multiplying $\bfq$ were replaced by a constant \eq{2dft}
would be similar to the usual expression for the form
factor.  We
 clarify this comparison by expressing the wave function of \eq{wf}
and \eq{2dft}
 in transverse position space, with $\bfB$ canonically conjugate to
 the 
transverse  momentum variable $\bfkappa$:
\bea \psi(x,\bfB)=\frac{1}{\sqrt{x(1-x)}}\int {d^2\bfkappa \over (2\pi)^2}e^{i\bfkappa\cdot\bfB}\psi(x,\bfkappa),\\
={\sqrt{x(1-x)}\over 2\pi} g K_0(\sqrt{m_1^2(1-x)+m_2^2x-M^2x(1-x)}\;B),\label{wf1}
\eea
with the phase space factor $\frac{1}{\sqrt{x(1-x)}}$  incorporated
in the wave function. Then 
 the form factor \eq{2dft} can be re-expressed as
\bea F(Q^2)={1\over 2(2\pi)^3}\int_0^1 dx\int
d^2\bfB|\psi(x,B)|^2e^{-i\bfq\cdot(1-x)\bfB}.
\label{coord}\eea
Further simplify by
replacing the relative transverse 
 position variable $\bfB$ by the transverse position variable of the charged parton $\bfb_1\equiv\bfb$. We have
 \bea
 \bfB=\bfb_1-\bfb_2\\
 0=\bfb_1(x)+\bfb_2(1-x),\\
 \bfB=\bfb/(1-x),\label{newv}\eea
 where the middle equation sets the transverse center of $P^+$ momentum to zero.
 Use \eq{newv} in \eq{coord} to find
\bea F(Q^2)={1\over 2(2\pi)^3}\int_0^1{dx\over (1-x)^2}\int d^2\bfb  |\psi(x,{\bfb\over1-x})|^2e^{-i\bfq\cdot\bfb}\label{coord1},\eea
which can be re-written as
\bea F(Q^2)={1\over (2\pi)^2} \int d^2\bfb  \rho(b)
  e^{-i\bfq\cdot\bfb}\label{coord2}\eea
with  the transverse density $\rho(b)$  given by
\bea \rho(b)={1\over4\pi}\int_0^1  {dx\over
  (1-x)^2}|\psi(x,{\bfb\over1-x})|^2={g^2\over2(2\pi)^3}\int_0^1  dx{x \over
  (1-x)}  K_0^2(\sqrt{m_1^2(1-x)+m_2^2x-M^2x(1-x)}\;{b\over1-x})\eea
The transverse density $\rho(b)$ has been derived previously
\cite{mbimpact,diehl2} as the integral of
the impact parameter generalized parton distributions GPD $\rho(x,b)$  
over all values of $x$. The
quantity $\rho(x,b)$ gives the probability that a quark of
longitudinal momentum fraction $x$ resides at a
transverse position \cite{mbimpact,diehl2}.
For the present model
\bea 
\rho(x,b)={g^2\over2(2\pi)^3}{x \over
  (1-x)}  K_0^2(\sqrt{m_1^2(1-x)+m_2^2x-M^2x(1-x)}\;{b\over1-x}).\label{gpd}\eea
The transverse density is also the integral of the
three-dimensional infinite momentum frame density $\rho(x^-,b)$ over
all values of the longitudinal position coordinate \cite{miller:2009qu}.

The transverse density is directly obtainable from
experiment via the inverse Fourier transform of \eq{coord2} 
 provided the electromagnetic form factor is measured for
sufficiently large
values of $Q^2$. 
The
momentum transfer is transverse in direction so that information about
the longitudinal position or momentum is not available. There is 
no way to use only measured values of $F(Q^2)$ 
 to determine 
$\rho(x,b)$.

\subsection{Singular central density}
Before proceeding it is  worthwhile to point out that the model
central density is singular. This arises as a  consequence of the zero
range nature of the $\Psi\phi\xi$ coupling. The transverse density
$\rho(b)$ is an integral involving the singular function $K_0(x)$
which varies as $\log{1\over x}$ for $x<<1$. 
The question of the singularity of the central transverse density $\rho(b)$ is
interesting to the present author because of recent work \cite{miller:2009qu}
showing that for the pion $\rho(b)$
is likely to approach infinity as $b$ approaches zero.
We may study the limit as $b$ approaches 0 by using the asymptotic
limit of the form factor \eq{simple}. The density for $b$ near  zero
is controlled by the form factor at large values of $Q^2$. We use the
inverse of 
\eq{coord1} to write
\bea
\lim_{b\rightarrow0}\rho(b)\sim \int_{Q_0}^{\epsilon/b} {dQ\over Q} \ln^2(Q)
,\eea
where $Q_0$ is a momentum transfer large enough so that so that
\eq{simple} is valid, and $\epsilon$ is a fixed positive number small
enough so that $J_0(\epsilon)=1$ to any desired  precision. Changing 
variables to $u=\ln Q$ shows that
\bea
 \lim_{b\rightarrow0}\rho(b)\sim \ln^3(b)/3
,\eea
which is the  central 
 singular charge density  arising from the $\log^2Q^2/Q^2$ behavior of
 the asymptotic form factor. 
\subsection{Transverse Charge Density from a more general perspective}

For a spin-0 system, the  form factor $F(Q^2)$, \eq{def} may be
computed, in the Drell-Yan DY  frame $(q^+=0,Q^2>0=\bfq^2)$,  by using 
\bea F(Q^2)={\langle p'|J^+(0)|p\rangle\over 2p^+}.\eea\label{fdef}
The spatial structure of a  nucleon can be examined if
one  uses \cite{soper1,mbimpact,diehl2}.
The state with transverse center of mass
$\bfR$ set to 0 is formed by taking a  linear superposition of
states of transverse momentum:
\be
\left|p^+,{\bf R}= {\bf 0},
\right\rangle
\equiv {\cal N}\int \frac{d^2{\bf p}}{(2\pi)^2} 
\left|p^+,{\bf p},  \right\rangle,
\label{eq:loc}
\ee
where $\left|p^+,{\bf p}, \lambda \right\rangle$
are plane wave  states
and
${\cal N}$ is a normalization factor satisfying
$\left|{\cal N}\right|^2\int \frac{d^2{\bf p}_\perp}{(2\pi)^2}=1$.
The normalization of the states is given by 
\bea\langle {p'}^+,\bfp'| {p}^+,\bfp\rangle
=2p^+(2\pi)^3  \delta({p'}^+-p^+)\delta^{(2)}({\bfp}'-\bfp).\label{norm}\eea
References~\cite{mb1,diehl} use  
wave packet treatments that  avoid states 
normalized to $\delta$ functions, but  this  leads to the
same results as using \eq{eq:loc}. Note however,  the relevant range of integration
in \eq{eq:loc} must be restricted to $|\bfp|\ll p^+$ to maintain the interpretation
of a nucleon moving with well-defined longitudinal momentum\cite{mb1}. Thus we use
a frame with very large $p^+$. It is in just such a frame that 
the interpretation of a nucleon as a  set of a large number of partons is valid.  


We evaluate 
the density operator in the infinite momentum frame in which the
spatial coordinates are $x^-=(t-z)/\sqrt{2},\bfb$,
and time is $x^+=(t+z)/\sqrt{2}=0$. We therefore do not write the $x^+$ dependence in any function below.
The infinite momentum frame charge density operator (in units of the
proton charge) is given by  
\bea \hat{\rho}_\infty(x^-,\bfb)\equiv J^+(x^-,\bfb)=\phi \stackrel {\leftrightarrow} {\partial^+}\phi
(x^-,\bfb),\label{imfop}\eea
and the density itself by
\bea {\rho}_\infty(x^-,\bfb)={ \left\langle p^+,{\bf R}= {\bf 0}
\right| \hat{\rho}_\infty(x^-,\bfb)
\left|p^+,{\bf R}= {\bf 0}
\right\rangle
\over 
\left\langle p^+,{\bf R}= {\bf 0}
|p^+,{\bf R}= {\bf 0}
\right\rangle}.\label{cdnes}\eea

Use translational invariance in the form 
$
\hat{\rho}_\infty(x^-,\bfb)=e^{i\hat{p}^+x^-}
e^{-i\bfp\cdot\bfb}\hat{\rho}_\infty(0)e^{+i\bfp\cdot\bfb}e^{-i\hat{p}^+x^-}$ along
with \eq{cdnes}, \eq{eq:loc} and \eq{def} to determine that  
\bea
\int dx^-\rho_\infty(x^-,\bfb)= \frac{1}{(2\pi)^2}\int d^2\bfq
F(Q^2=\bfq^2)e^{-i\bfq\cdot\bfb}=\rho(b)
.\eea
Thus one recovers the two-dimensional Fourier transform of \eq{coord2}.

\subsection{Mean-squared transverse radii and mean-squared effective
  radii}
The two-dimensional  Fourier transform of \eq{coord2} may be expanded
as a power series in $Q^2$ as
\bea
\lim_{Q^2\rightarrow0}F(Q^2)=1-{Q^2\over 4} \langle b^2\rangle,\label{true}\eea
where the mean-squared transverse radius $ \langle b^2\rangle $ is
given in terms of the transverse density as 
\bea  \langle b^2\rangle=\int d^2b b^2 \rho(b),\eea
and a direct relation with the transverse density is evident.
In contrast, the usual procedure is to write 
\bea \lim_{Q^2\rightarrow0}F(Q^2)=1-{Q^2\over 6}{R^*}^2,\label{effect}\eea
where we denote ${R^*}^2$ the effective mean-squared  radius\cite{Miller:2007kt} .
The quantity ${R^*}^2$ has no direct relationship with a density
unless the system is non-relativistic. Thus we maintain that  $
\langle b^2\rangle$ is the basic quantity related to an underlying
density.
However, once the effective mean-squared  radius ${R^*}^2$ is
determined, the  fundamental $\langle b^2\rangle$ is known immediately
because a comparison of \eq{true} and \eq{effect} reveals that
\bea 
 \langle b^2\rangle={2\over3}{R^*}^2.\eea

 \section{Wave function as a function of three position variables}
The previous section shows that the form factor is simply related to
the three-dimensional coordinate-space density that depends on $(x^-,\bfb)$ 
in the infinite momentum frame. Given the simplicity of our model, we
should be able to identify a wave function and density.

The basic idea is that the position variable of a particle $x^-$ is canonically conjugate to the
plus-component of the momentum. The momentum of the charged
constituent is $k^+=x P^+$, and its canonical longitudinal position variable is
$x^-$ with $[x^-,k^+]=i=[x^-,x]P^+$ \cite{Pirner:2009zz}.
 The canonical longitudinal
position variable for the other particle can be taken as $-x^-$. 
So we can convert the wave function
$\psi(x,\bfB)$ of \eq{wf1} to one expressed entirely in   coordinate
space. We find
\bea \psi(x^-,\bfB)=\sqrt{P^+\over 2\pi}\int_0^1\;dx\;\psi(x,\bfB)e^{ixP^+x^-},\label{coord22}\eea
which preserves the normalization condition that $F(Q^2=0)=1$. Note
that $\bfB$ is a relative variable and that $x^-$ is the variable for
the position of the charged constituent.   This
wave function displays no spherical symmetry--the longitudinal $x^-$
and transverse position dependence $\bfB$ are not related.
Another point is that the wave function \eq{coord22} explicitly depends
on the momentum $P^+$. As $P^+$ approaches infinity,  the value  
of $x^-$ must be very small to prevent  $x^-P^+$ from being
very large and causing the integral on \eq{coord22} 
 to vanish. This means  that the system can be thought of as
having a pancake or disc shape. For this reason, the position $b=0$
really does correspond to the center of the hadron.  We shall show below, that in the
non-relativistic limit, rotational symmetry emerges.

The wave function can be computed in closed form for the special case
$M=0,m_1=m_2=m$. Using \eq{wf1} in \eq{coord22} with the stated
parameters leads to the result
\bea\psi(x^-,\bfB)=\sqrt{P^+\over 2\pi}e^{i{1\over2}P^+x^-}g\;
K_0({m}\;B)\int_0^1\;dx\;{\sqrt{x(1-x)}\over 2\pi} e^{ixP^+x^-}\\
= \sqrt{P^+\over 2\pi}e^{i{1\over2}P^+x^-}g\;K_0({m}\;B)\frac{\pi
  J_1\left(\frac{P^+x^-}{2}\right)}{4P^+x^- }\label{simp}\eea
For this simple example, the $x^-$ and $\bfB$ dependence factorizes,
showing the explicit violation of rotational symmetry. The formula
\eq{simp} shows also how the spatial extent    contracts with the
increase in value $P^+$.
We note that it is not useful to use  the spatial wave function  to compute  the
form factor because of the appearance of the factor $1-x$ in the
exponential
of \eq{coord}.

\section{The Rest Frame Charge Distribution is Generally Not
  Observable}
The concept of a charge density that depends on three spatial
variables, but not on the time, is inherently non-relativistic. This
is because the use of only three variables involves 
 replacing a four-dimensional quantity by one involving only three
dimensions.
One procedure, discussed above in Sect.~II, is to evaluate the   Feynman diagram of
Fig.~\ref{ff} by integrating over the $k^-$  component of the virtual
momentum $k$. This  leads to a formalism in which the form factor
depends on a Fourier transform of the square of a wave function that
depends both on position $\bfB$ and momentum $x$ variables.

\begin{figure}
\unitlength.8cm
\begin{picture}(14,8.2)(1.5,.6)
\includegraphics{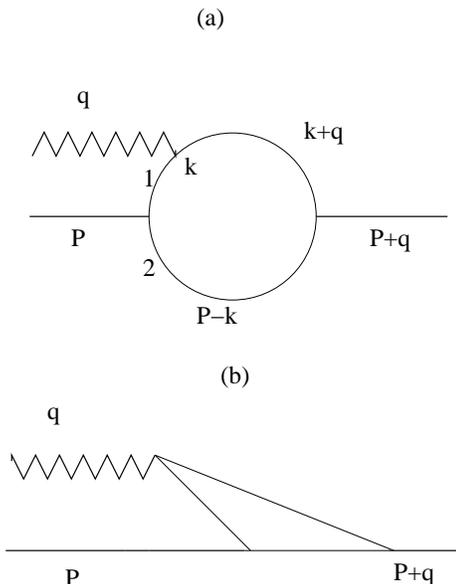}
\end{picture}
\caption{\label{topt}
Two TOPT diagrams for the form factor with the photon coupling
  to the particle of mass $m_1$. }
\end{figure}

One can try to recover the more familiar 
three-spatial dimension formalism by evaluating  
 the Feynman diagram of
Fig.~\ref{ff}  using  the time-ordered-perturbation theory  TOPT
formalism in the rest frame. One proceeds by integrating over all times, with the
exponential oscillating factors converted into energy denominators. 
In  the   TOPT formalism  any
given Feynman diagram is the sum of several TOPT diagrams. In the
present case, the sum of the two TOPT diagrams of Fig.~\ref{topt} leads
to the Feynman diagram of Fig.~\ref{ff}. Only Fig.~\ref{topt}a
corresponds to measuring a density. The term of Fig.~\ref{topt}b
corresponds to the hadronic part of the incident photon wave function
interacting with the target. 

\begin{figure}
\unitlength1cm
\begin{picture}(14,8.2)(-2.,-.6)
\includegraphics{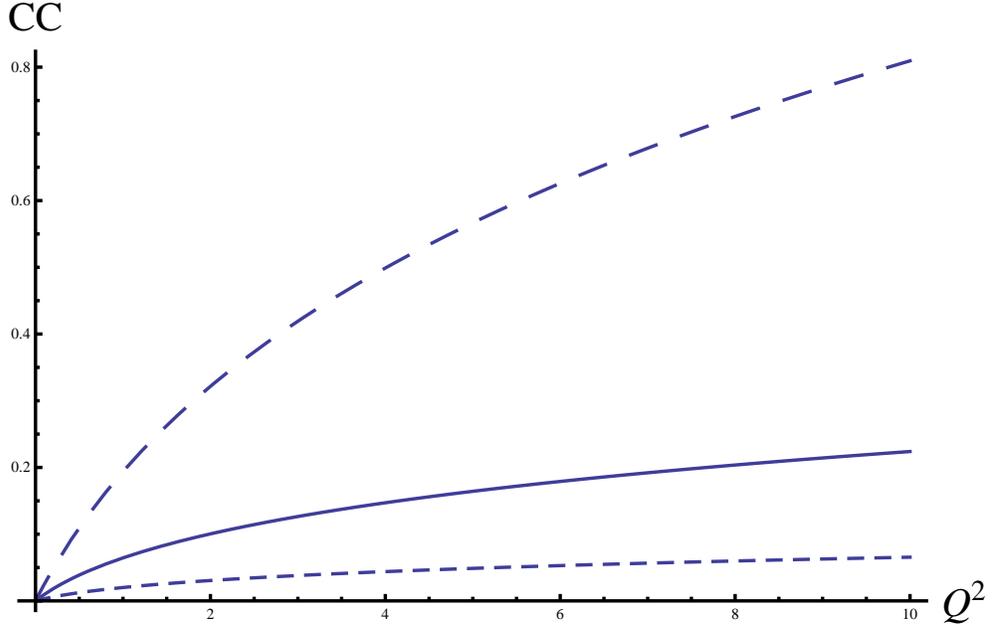}
\end{picture}
\caption{\label{cc} Non-conservation of current is measured by the
  deviation of $CC$ from 0 as a function of $\bfq^2=Q^2$ (in units of
  $M^2$). 
Solid curve
  $m=0.51\;B/M=0.04$, short-dashed curve $m=0.501\;B/M=0.004$ long-dashed curve $m0.6\;B/M=0.33$
( $m$in units of $M$)} 
\end{figure}

One can examine the contribution of the term of Fig.~\ref{topt}a, $F_{2a}(Q^2)$. It is given
\cite{Gunion:1973ex}
by 
\bea \langle P+q|J^\mu(0)|P\rangle\rightarrow(2P+q)^\mu F_{2a}(Q^2)={g^2\over (2\pi)^3}\int{d^3p\over 2E_1{2E_1}'2E_2}
{(p^\mu_1+{p'}^\mu_1)\over (E_P-E_1-E_2)(E_{\bfP+\bfq}-{E'}_1-E_2)},\label{g1}\eea
an expression that leads to the correct result in the infinite
momentum frame ($P\rightarrow\infty$) \cite{Gunion:1973ex}. This
expression can be interpreted as involving an initial and a final
state wave function if one interprets the energy denominators
(multiplied by 
phase space factors)  as wave
function expressed in momentum space. 
 The symbol $\rightarrow$ used above refers
to the approximation of keeping only a single TOPT diagram. For simplicity we take the
example, $m_1=m_2=m$ and also  work in the target rest frame:
$\bfP=\bf{0}$ to isolate the rest frame charge distribution.
Then
$E_P=\sqrt{P^2+M^2}=M,\;E_{\bfP+\bfq}=\sqrt{(\bfP+\bfq)^2+M^2}=\sqrt{\bfq^2+M^2},\;
E_1=\sqrt{p^2+m^2},
\;{E'}_1=\sqrt{(\bfp+\bfq)^2+m^2}$.
The four vector $(p^\mu_1+{p'}^\mu_1)=[E_1+{E'}_1,2\bfp+\bfq]$. There
are three integrals appearing on the right-hand-side of \eq{g1}:
\bea && I_1(\bfq^2)\equiv \int{d^3p\over 2E_1{2E_1}'2E_2}
{(\sqrt{p^2+m^2}+\sqrt{(\bfp+\bfq)^2+m^2})\over (E_P-E_1-E_2)(E_{\bfP+\bfq}-{E'}_1-E_2)}\\
&&\hat{\bfq}J_2(\bfq^2)\equiv\int{d^3p\over 2E_1{2E_1}'2E_2} {2\bfp\over (E_P-E_1-E_2)(E_{\bfP+\bfq}-{E'}_1-E_2)}\\
&&\hat{\bfq}J_3(\bfq^2)\equiv\int{d^3p\over 2E_1{2E_1}'2E_2}{\bfq\over
  (E_P-E_1-E_2)(E_{\bfP+\bfq}-{E'}_1-E_2)}.\eea
Thus we arrive at  the four-vector equality
\bea  (2P+q)^\mu F_{2a}(Q^2)={g^2\over
  (2\pi)^3}[I_1,\hat{\bfq}(J_2+J_3)],\label{g2}\eea
Maintaining current conservation requires that the matrix element of
$q_\mu J^\mu$ vanishes. Taking the scalar product of \eq{g2} with
$q_\mu$ leads to the requirement:
\bea
0=q^0I_1-|\bfq|(J_2+J_3)\equiv CC\;I_1(\bfq^2=0),\label{g3}\eea
where $q_0=\sqrt{\bfq^2+M^2}-M$. The right-hand-side of \eq{g3} is
defined as $CC\;I_1(0)$ so that comparing $CC$ to unity provides a
reasonable measure of  the failure of this approximation to uphold
current conservation. We   express all momenta and mass in units  of the
target mass (=1), take as an examples $m=0.501,\;0.51$ and 0.6 and plot the numerical
results in Fig.~\ref{cc}. Both $CC$ and
  $\bfq^2=Q^2$ are measured in units of the target mass $M$, taken as
unity. Thus the natural scale of any quantity is unity. We see that
current conservation is massively violated in the rest frame for
systems in which $B/m=(2m-M)/m$ is not very small. In that case, 
 the expression that potentially depends on the square of the wave
function or density has no independent physical reality. 

\section{Neutral Systems}
Previous work \cite{Miller:2007uy} showed that the 
central transverse density of the neutron is
negative. This contrasts with the long held view that there must be
positive charge density at the center to neutralize the effects  of
  a negatively charged pionic   cloud  that occupies the
  exterior. This result demands interpretation
  \cite{Miller:2008jc}-
\cite{Rinehimer:2009sz}.

One relevant question is whether or not the intuition that a neutral
system consisting of a heavy charge positively charged particle and a
negatively
charged lighter particle disobeys the standard intuition that the
averaged
squared charged radius is negative, when the charge density is 
evaluated in the infinite momentum frame. We examine this question in
our
model by taking the $\phi$ (of mass $m_1$) to be positively charged
and the $\xi$ (of mass $m_2<m_1$ to
be negatively charged.

The form factor of this model can be obtained 
by using  \eq{final}  
by including  a second term obtained by interchanging $m_1$ and $m_2$ and
putting a minus sign in front. That operation gives the result
\bea F(Q^2)={g^2\over 4\pi^2}\int_0^1dx x[
{ {\rm Tanh}^{-1}[{\sqrt{Q^2}(1-x)\over\sqrt{4x\;
	m_2^2+4v(1-x)m_1^2-x(1-x)M^2+(1-x)^2Q^2}}]\over\sqrt{Q^2}\sqrt{4x\; m_2^2+4(1-x)m_1^2-x(1-x)M^2+(1-x)^2Q^2}}-\nonumber\\{ {\rm Tanh}^{-1}[{\sqrt{Q^2}(1-x)\over\sqrt{4x\; m_1^2+4(1-x)m_2^2 -x(1-x)M^2+(1-x)^2Q^2}}
]\over\sqrt{Q^2}\sqrt{4x\; m_2^2+4(1-x)m_1^2-x(1-x)M^2+(1-x)^2Q^2}}].
\label{final1} \eea

\begin{figure}
\unitlength1cm
\begin{picture}(14,8.2)(-3.5,-.6)
\includegraphics{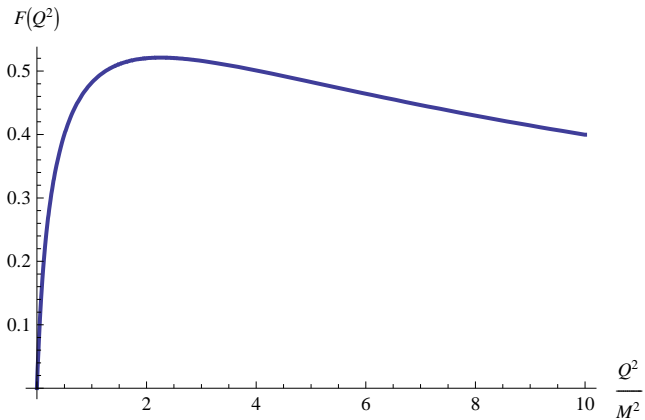}
\end{picture}
\caption{\label{neutral} Form factor for a neutral system with one
  heavy $m_2=M$ and one light $m_1=0.14\; M$ negatively charged constituent.  } 
\end{figure}

The results of a numerical evaluation using $m_1=M$ and $m_2=0.14M$
are shown in Fig.~\ref{neutral}. One observes  the rise of $F(Q^2)$
from zero, which is the effect expected from non-relativistic,
rest-frame considerations. The effective squared radius, defined in
\eq{effect} is indeed negative. This is the same as expected from the
intuition that 
the negatively charged light particle
resides on the outside edge of  the system.

\begin{figure}
\includegraphics[width=8cm]{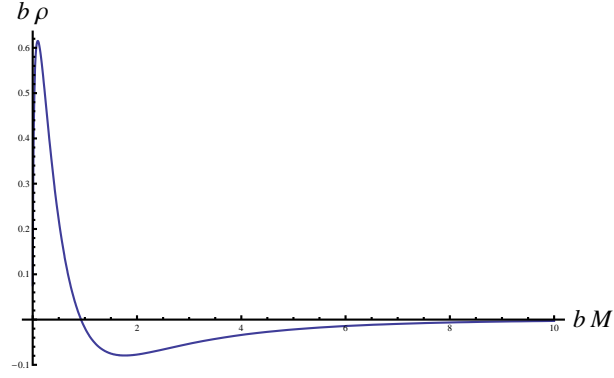}
\caption{\label{nrhob} Transverse charge density for a neutral system
  of
a positively charged heavy object and a negatively charged lighter object}
\end{figure}

One obtains the analytic result for the charge radius 
 by taking the  limit of very low $Q^2$ 
 in the expression for the form factor \eq{final1}. The expression is
 simplified if one uses the (relevant for nucleon)\cite{assum}
  case of $m_1=M$. 
Then  one finds:
\bea  M^2{R^*}^2=-{g^2\over 96\pi^2}
\left(
\frac{\left(\frac{m_2^2}{M^2}-1\right)
   \left
(-2 \tan ^{-1}\left(\frac{\sqrt{{m_2^2\over M^2}
   \left(4-\frac{m_2^2}{M^2}\right)}}{m_2^2}
\right
)
   \sqrt{\frac{m_2^2}{M^2}} \left(\frac{m_2^2}{M^2}-5\right)-2
   \sqrt{4-\frac{m_2^2}{M^2}}\right)}{m_2^2
   \left(4-\frac{m_2^2}{M^2}\right){}^{3/2}}-\log
   \left(\frac{m_2^2}{M^2}\right)\right)
.\eea
A very accurate approximation (better than 1\% for $m^2_2\le0.14M^2$)
is \bea M^2{R^*}^2=-{g^2\over 4\pi^2}\left(
\frac{M^2}{48 m_2^2}+\frac{1}{96} \left(1-4 \log
   \left(\frac{m_2^2}{M^2}\right)\right)+\frac{11}{512} \pi 
   \sqrt{\frac{m_2^2}{M^2}}-\frac{5 \pi }{192
   \sqrt{\frac{m_2^2}{M^2}}}-\frac{7 m_2^2}{288 M^2}\right).\eea
The radius is dominated by a singular term proportional to $1/m_2^2$.
Thus as expected the lighter constituent drifts to the edge of the
nucleon. The conventional expectation is borne out on the light front.
This is shown in  more detail by plotting $b\rho(b)$, as shown in
Fig.~\ref{nrhob}.
 The positive
charge density  is concentrated at the center and the negative at the
edge. This finding does not contradict the explanations offered in 
Refs.~\cite{Miller:2008jc}-
\cite{Rinehimer:2009sz}. Ref.\cite{Miller:2008jc} argues that 
negative charge at high $x$ corresponds to negative charge at small
values of $b$. The $N\pi$ model of Ref.~\cite{Rinehimer:2009sz} shows
that one must include 
the finite size of the nucleon to obtain a computed $F_1$ that looks
like the measured function. Thus the point-like nature of the
constituents used here is unsurprisingly not  relatistic. Moreover,
in that model negatively charged pions reside both at the edge and at
the center of the nucleon. The implication of \cite{Miller:2008jc}
 is that the pions may  have large values of longitudinal momentum
 fraction. 
This expectation is borne out by the model calculation
\cite{Strikman:2009bd}. Thus in pion cloud models of the nucleon pions
that have a large longitudinal momentum tend to reside near the center
of the nucleon.

\section{Non-relativistic Limit}
The conventional lore is that the electromagnetic form factor is the
Fourier transform of the charge density. In this section we see how
this idea emerges by taking the non-relativistic limit.

Our starting point is the wave function \eq{wf} and the form factor
\eq{2dft}. 
Recall that the quantity
$x=k^+/P^+$. We work in the rest frame  and take the non-relativistic
limit in which the energy $k^0=m_1$, and $k^+=m_1+\kappa^3$, where
$\kappa^3$ is the third-component of  the relative longitudinal momentum.
  Then \cite{Brodsky:1989pv,Frankfurt:1981mk} 
\bea
x=\frac{m_1+\kappa^3}{M},\quad 1-x=\frac{M-m_1-\kappa^3}{M}=\frac{m_2-B-\kappa^3}{M},\label{a1}\eea 
 where  in 
conformation with non-relativistic notation, we define the positive
binding energy $B$ so that 
\bea M\equiv m_1+m_2-B.\label{bdef}\eea
To obtain the non-relativistic expression we express
 the denominator appearing in \eq{wf} in terms of $\kappa^3$. This gives
\bea&& M^2-{\bfkappa^2+m_1^2\over x}- {\bfkappa^2+m_2^2\over 1-x}\label{rel23}\\
&&=M^2-M[{\bfkappa^2+m_1^2\over m_1+\kappa^3}+{\bfkappa^2+m_2^2\over m_2-B-\kappa^3}]\\&&
\approx
M^2-M[{\bfkappa^2+m_1^2\over m_1}\left(1-{\kappa^3\over    m_1}+({\kappa^3\over m_1})^2\right)
+{\bfkappa^2+m_2^2\over m_2}\left(1+{\kappa^3+B\over
  m_2}+({\kappa^3+B\over m_2})^2\right)]\\&&
\approx(\bfkappa^2+\kappa_3^2)({1\over m_1}+{1\over m_2})    +m_1+m_2+B]\\&&
=2M(-B-{\kappa^2\over 2\mu}),\label{prop12}\eea
where \bea 
{\kappa}^2\equiv\bfkappa^2+\kappa_3^2,\; 
\vec{\kappa}=\bfkappa
+\kappa^3 \hat{\bfz}, \;
\mu\equiv{m_1m_2\over m_1+m_2}.\eea
In going from \eq{rel23} to \eq{prop12} 
we have ignored terms in $v/c=k/m$ of order three and higher. The
result is that
    \eq{prop12} is
recognizable as $2M$ times the inverse of the non-relativistic propagator.

The next step is to determine the coordinate form of the
non-relativistic wave function $\psi_{NR}(\vec{r})$ (where $\vec{r}$
is canonically conjugate to $\vec{\kappa}$)
 and to show that the non-relativistic
form factor is a three-dimensional Fourier transform of 
$\left|\psi_{NR}(\vec{r})\right|^2$. First use the non-relativistic
approximation \eq{prop12} in \eq{wf} to find
\bea \psi_{NR}(\vec{\kappa})={-2\mu g\over \kappa^2+\lambda^2},\;
\lambda^2\equiv 2\mu B.\label{nrp}\eea
The coordinate-space wave function  $\psi_{NR}(\vec{r})$ is given by
\bea 
\psi_{NR}(\vec{r})={1\over (2\pi)^{3/2}}\int d^3\kappa 
e^{i\vec{\kappa}\cdot\vec{r}}\psi_{NR}(\vec{\kappa})=-{\mu g\over
  M}\sqrt{\pi\over 2}{e^{-\lambda r}\over r}.\label{nrr}\eea
The expression \eq{nrr} is seen as the standard result obtained for
the bound state of a two-particle system interacting via an attractive
delta function potential.

The wave functions \eq{nrp} and \eq{nrr} 
enable us to examine the condition needed for the approximations
\eq{a1} to be valid. For \eq{a1} to work we need $\kappa^2\ll
m_{1,2}^2$, but from the wave functions $\kappa^2\sim \lambda^2$ so
that we require 
\bea {\mu B\over m_{1,2}^2}\ll 1\label{cond0}\eea
for the non-relativistic approximation to be valid. More specifically
let ${\cal M}$  be the lighter of $m_1,m_2$, then we may write
 the approximate condition as
\bea {B\over {\cal M}}<<1.\label{cond}\eea

The non-relativistic form factor $F_{NR}(Q^2)$ is obtained by using
\eq{nrp} in the expression for the form factor \eq{2dft}, and taking
the non-relativistic limit defined by the expressions:
\bea && dx\rightarrow {d\kappa^3\over (m_1+m_2)}\\&&
x(1-x)\rightarrow {m_1m_2\over m_1+m_2}\\
&& (1-x)\bfq\rightarrow {m_2\over m_1+m_2}\bfq.\eea
The result is
\bea F_{NR}(Q^2)={1\over 2(2\pi)^3\mu}\int d^3
r\left|\psi_{NR}(\vec{r})\right|^2e^{-i \bfq\cdot \bfr {m_2\over m_1+m_2}}.\eea
This is the usual expectation that the form factor is  a 
three-dimensional Fourier transform
of the wave function. We may evaluate the integral immediately to find
\bea
F_{NR}(Q^2)={\tan^{-1}{Q m_2\over 2(m_1+m_2)\lambda}
\over {Qm_2\over 2(m_1+m_2)\lambda}},\label{fnrgen}
\eea
where $Q=|\bfq|$ and 
the coupling constants and other constants enter in such a manner as
to make $F_{NR}(Q^2=0)=1$.

 In
 the remainder of this section we study the accuracy of the non-relativistic approximation by
comparing the results of  using \eq{fnrgen} with the model-exact
results of using \eq{final} for
several examples.

\subsection{ Bound state of two equal mass particles}
With equal masses $m_1=m_2=m$ the bound state can be thought of as a
toy meson
or a deuteron. Use $m_1=m_2=m$ in \eq{nrp} leads to the 
%
non-relativistic wave function
$\psi_{NR}^{(2)}(\kappa)$ with
\bea \psi_{NR}^{(2)}(\kappa) ={g\over 2M( \kappa^2+\lambda_2^2)},\label{psinr}\eea
with 
\bea \lambda_2^2=m B.\eea T
he 
 coordinate space wave function $\psi_{NR}(r)$ is then 
\bea \psi_{NR}^{(2)}(r)=\sqrt{\pi\over2}{g\over 2M} {e^{-\lambda_2 r}\over r}.\eea
 Thus the wave
function is the usual bound state wave function one obtains with a
delta function binding interaction. We obtain the non-relativistic
version of the form factor by 
using  $m_1=m_2=m, \lambda\rightarrow\lambda_2$ in \eq{fnrgen} to find
\bea
F_{NR}^{(2)}(Q^2)={\tan^{-1}{Q\over 4\lambda_2}\over {Q\over 4\lambda_2}},\label{fnr}
\eea
where $Q=|\bfq|$ and 
the coupling constants and other constants enter in such a manner as
to make $F_{NR}^{(2)}(Q^2=0)=1$.

\begin{figure}
\unitlength.91cm
\begin{picture}(16,10.2)(-3.5,-1)
\includegraphics{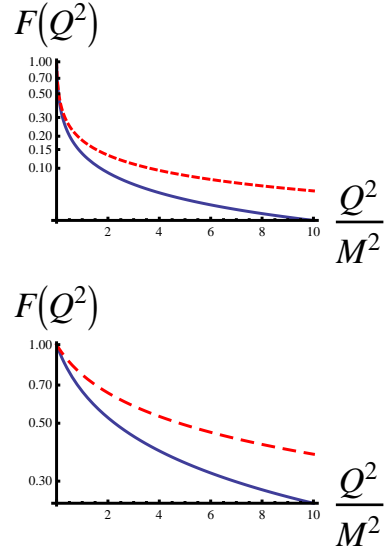}
\end{picture}
\caption{\label{nonrel} Exact \textit{vs} non-relativistic form
  factors for the case $m_1=m_2=m$. Solid curve- exact result, dashed
  curve-non-relativistic  limit. Upper panel deuterium-like
  kinematics in which $B=0.002 M$. Lower panel- $B=0.1M$.
} 
\end{figure}
We study the non-relativistic approximation numerically by comparing the exact
model results
\eq{2dft} with those of the non-relativistic approximation
\eq{fnr}. See Fig.~\ref{nonrel}.
The figure  shows two sets of results. In the upper panel the binding energy $B=0.002\;M$. This corresponds roughly to deuteron kinematics, in which the binding energy is of the order of a 0.004 of the deuteron mass. We see that the non-relativistic approximation is not accurate for values of $Q^2/M^2$ greater than about 1. If one increases the binding energy to 0.1 $M$,
one sees that the non-relativistic approximation is not accurate for any value of $Q^2$. If one approximates a nucleon by taking $M=$ 1 GeV, then
$m=0.55$ GeV, which is much larger than a $u,d$ constituent quark
mass. Thus the range of masses for which the non-relativistic
approximation is valid is very narrow indeed.

We can gain some insight into the nature of the relativistic corrections to the charge radius by studying the low $Q^2$ limit of the form factor of \eq{final}.  One finds
\bea \lim_{Q^2\rightarrow0}F(Q^2)= 1-{Q^2{R^*}^2\over 6},\label{effects}\eea
where we use the notation ${R^*}^2$ to denote an effective 
radius squared that is not generally associated with the expectation
of the square of a radius operator weighted by a density. The explicit
evaluation gives
\bea M^2{R^*}^2=
\frac{\left(\frac{1}{\gamma ^3}+48 \gamma \right) \cot ^{-1}(2 \gamma )+\frac{2}{\gamma ^2}-24}{16 \left(\left(2 \gamma +\frac{1}{2 \gamma }\right) \cot
   ^{-1}(2 \gamma )-1\right)},\eea
   and
   \bea \gamma^2\equiv{m^2\over M^2}-{1\over4}={B\over 2M}+{B^2\over 4M^2}.\eea

\begin{figure}
\unitlength.91cm
\begin{picture}(10,8.2)(-.5,.0)
\includegraphics{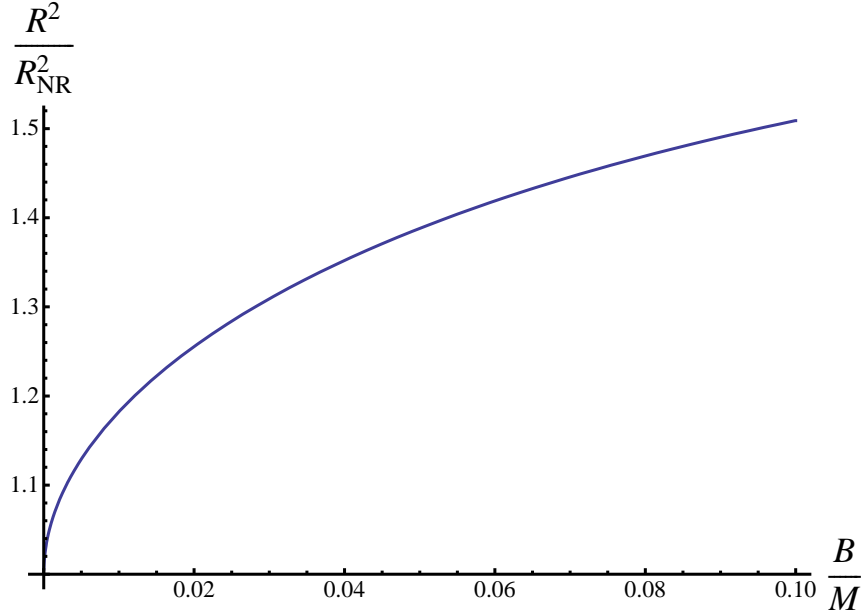}
\end{picture}
\caption{\label{Rsq} Ratio of exact 
to non-relativistic effective square radii  for the
  case
 $m_1=m_2=m$ as a function of the ratio of the binding energy $B$ to
  the hadronic mass $M$. This is also the ratio of the true value of
  $\langle b^2\rangle$ to its non-relativisitic version. } 
\end{figure}

   The non-relativistic limit corresponds to the limit of small values
   of $\gamma$, which corresponds to a small value of $B/M$. So we expand the previous result to order $B/M$ to find
   \bea
   M^2{R^*}^2\approx\frac{\left(12288-2816 \pi ^2+195 \pi ^4\right) B}{48 M \pi ^4}+\frac{\sqrt{\frac{B}{M}} \left(128 \sqrt{2}-25 \sqrt{2} \pi ^2\right)}{4 \pi
   ^3}+\frac{64-5 \pi ^2}{8 \pi ^2}+\frac{\sqrt{2}}{\sqrt{\frac{B}{M}} \pi }+\frac{M}{4 B}
\label{expand}\eea
   The non-relativistic value of the mean square radius, $R_{NR}^2$
   (which is a true mean square radius)
is obtained by expanding the form factor for small values of $Q^2$:
   \bea R^2_{NR}={1\over 8 m B}\approx {1\over 4MB},\label{rsqnr}\eea
   which corresponds to the leading term of \eq{expand} in the limit that $B $ approaches 0.
   Comparing \eq{expand} with \eq{rsqnr} shows that the former
   contains a  series of terms
   that represent the boost corrections to the non-relativistic result.
   Each correction is positive and can be substantial. 
 The figure \ref{Rsq} shows the ratio of the exact of the mean square
   radius  to the non-relativistic approximation  as a function of
   $B/M$. We see that the non-relativistic approximation works  well
   only for 
very small  values of $B/M$. 
Indeed,     the ratio of the leading correction  to the
   non-relativistic
result is given by
\bea  {{R^*}^2-R_{NR}^2\over R_{NR}^2}\approx {4\over \pi}\sqrt{2B\over
   M}.\label{eqm}\eea
For this ratio to be less than 10 \%, 
 ${B\over M}$ must  be less than one part in a thousand! Thus, within
   the framework of our  toy    model, the relativistic corrections
   can generally expected to be  very substantial.

\subsection{Quark-diquark model of the nucleon}

\begin{figure}
\unitlength.91cm
\begin{picture}(10,8.2)(-.5,.0)
\includegraphics{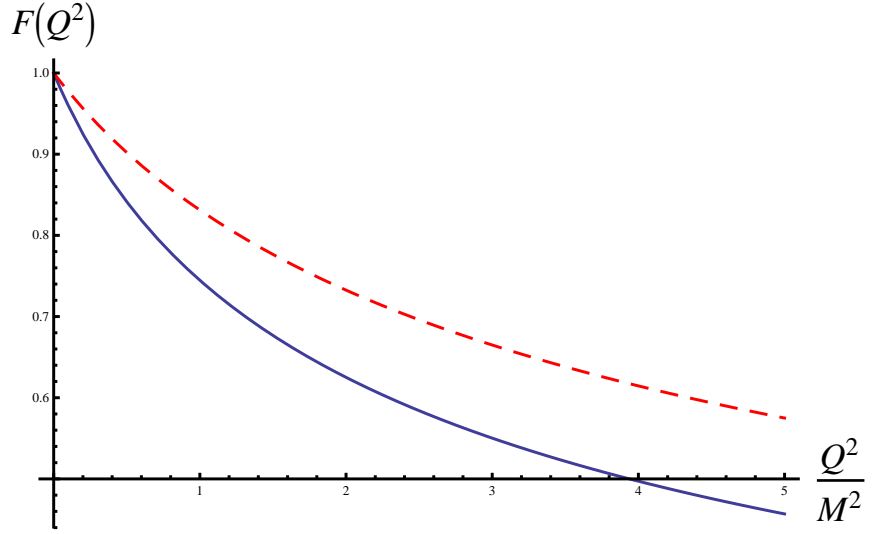}
\end{picture}
\caption{\label{qdiq} Exact \textit{vs} non-relativistic form factors 
for the case $m_2=2m_1, m=400$ MeV, $B=260\;{\rm  MeV}=0.276 \;M$. Solid
curve-exact, dashed non-relativistic.
} 
\end{figure}

Another interesting example is motivated by recent quark-diquark models
of the nucleon \cite{Oettel:2000jj,Horikawa:2005dh,Cloet:2008re}. 
We take $m_1=m, m_2=2m_1=2m$. Then from \eq{bdef} we have
$M=3m-B$. In these, models current quarks acquire a large constituent
mass
due to the effects of dynamical chiral symmetry breaking. Therefore we
take
$m=400$ MeV, and $M=940$ MeV which corresponds via \eq{bdef} to $B=260
$ MeV and $B/M$ = 0.276.
The non-relativistic expression for the form factor, $F_{NR}^{q2q}(Q^2)$
 for this case
is obtain 
using
the appropriate reduced mass as

 \bea
&&F_{NR}^{q2q}(Q^2)={\tan^{-1}{Q\over 3\lambda_{12}}\over {Q\over
    3\lambda_{12}}},\label{fnrq2q}\\&&
\lambda_{12}^2\equiv {4\over 3} mB.
\eea
Results comparing the exact form factor computed from \eq{final} 
with that of \eq{fnrq2q}
are shown in Fig.~\ref{qdiq}. The non-relativistic version
gives a poor approximation to the exact form factor for all values of
$Q^2$. This can be understood by considering the effective squared
radius 
${R^*}^2_{q2q}$ for this case. We find
\bea  {{R^*}^2_{q2q}-R_{NR}^2\over R_{NR}^2}\approx {6-\ln 2\over \pi}
\sqrt{B\over
   M}.\eea
The right-hand-side is evaluated as 0.887 for the present case, so
that there is a substantial relativistic correction to the quantity
$F_{NR}^{(2)}-1$ 
for any non-zero value of $Q^2$. This means that one can not take a
three-dimensional Fourier transform of the form factor to get a charge
density even if the constituent masses are large.

\subsection{Nuclear Physics and $m_1\ne m_2$}
\begin{figure}
\unitlength.91cm
\begin{picture}(10,8.2)(-.5,.0)
\includegraphics{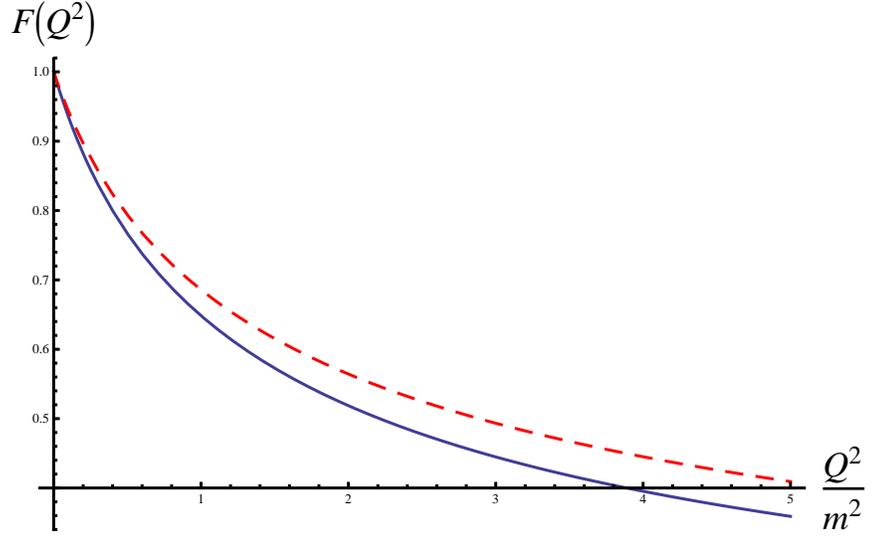}
\end{picture}
\caption{\label{A4} Exact \textit{vs} (solid curve) non-relativistic form
  factors (dashed curve) 
for $A=4$  for $Sm=46 $ MeV.
} 
\end{figure}

\begin{figure}
\unitlength.91cm
\begin{picture}(10,8.2)(-.5,.0)
\includegraphics{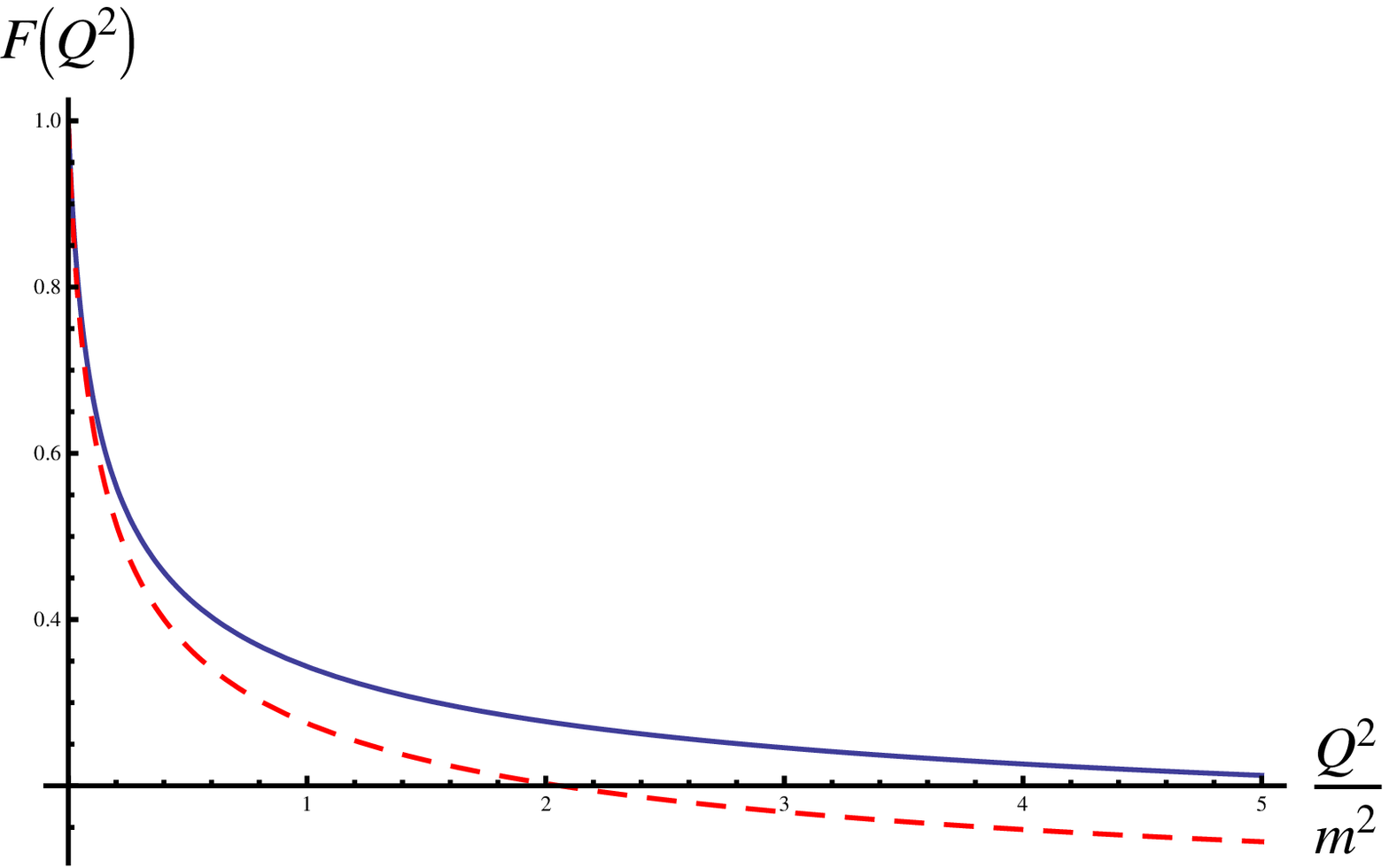}
\end{picture}
\caption{\label{A208} Exact (solid curve) \textit{vs} (dashed curve) 
non-relativistic form factors 
for $A=208$  for $Sm=46 $ MeV.
} 
\end{figure}

We consider masses that correspond to electron scattering from a
charged nucleon of mass $m$ (which is the free nucleon mass minus the
average binding energy per nucleon of 8 MeV)
bound in a nucleus of mass $M=m A$, with a
spectator  system of mass $m_2=(A-1)m +S $, where $S$ is the orbital
separation energy.  We measure all momenta in
terms of  $m=932 $ MeV, and take the  separation energy $S=0.05$ or
$Sm$ about
46 MeV.
 The results for $A=4$ and  $A=208$ are shown in Figs. \ref{A4} and \ref{A208}
The startling finding is that the relativistic effects reduce the form
factor for light nuclei, but increase it for heavy
nuclei. Furthermore, the relativistic effects are larger for heavy
nuclei than for light nuclei (for a fixed value of $S$.)

\begin{figure}
\unitlength.41cm
\includegraphics[width=8cm]{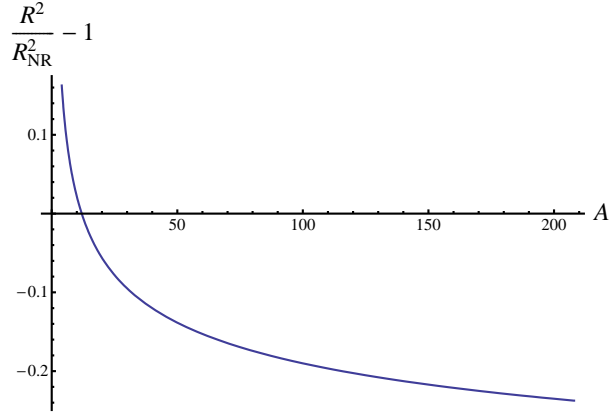}
\caption{\label{rsqa} Exact \textit{vs} non-relativistic effective
 radii,
$\frac{{R^*}^2}{{R_{NR}^*}^2}-1$  as a
  function of $A$  
  for $Sm=46 $ MeV. This is also the ratio of $\langle b^2\rangle$ to
 its non-relativistic counterpart.
} 
\end{figure}

We obtain some analytic understanding by expanding the effective
squared radius (defined in \eq{effects} in powers of $S$. We find 
\bea
m^2{R^*}^2=\frac{A-1}{4 A S}+\frac{\sqrt{(A-1) A} \left(4 A-(A-2) \log
  \left((A-1)^2\right)\right)}{4 \sqrt{2} A^2 \pi
  \sqrt{S}}+\cdots.\eea
We see that the first term is indeed the non-relativistic result, and
that the second term changes sign for the value  of $A$ that satisfies
the equation $4A-2(A-2)\ln(A-1)=0$ or $A\approx 12$. This
 displayed in Fig.~\ref{rsqa}. It is also seen that the relativity
causes very significant  effects on the effective radii. Except for
values of $A$ near 12, the changes are of the order of 10-15\%. I
expect that the
specific values shown in   Fig.~\ref{rsqa} are highly
model-dependent. Covariant models other than the $\Psi\phi\xi$ model used here
probably have  have effects of different sizes. However, the large
effects shown here cause one to wonder if relativity really may cause
the true
nuclear radii extracted from elastic electron scattering to differ by
10-20\% from those  appearing in tables. As noted above, we can expect
that the model employed here is a reasonable representation of the
lowest $s$-state of heavy nuclei for which the range of the
binding interactions is much less than the size of the system as a
whole. For such states, the results of Fig.~\ref{rsqa} should  be a
reasonably accurate guide, so that  significant effects of relativity
should be expected.

.

\section{Summary}
A relativistic model of a scalar particle $\Psi$ is a bound state of
two scalar particles $\phi$ and $\xi$ is used to elucidate
relativistic aspects of electromagnetic form factors.
 First, 
the form factor 
for the situation in which the $\Psi$ and $\phi$ carry a single unit of charge, but the
$\xi$ is neutral is computed using an exact  covariant calculation of
the 
lowest-order triangle diagram. This is followed by a  
another derivation using the  light-front technique 
of integrating over the minus-component of the virtual momentum in
Sect.~III 
that obtains the same form factor. This is also the result obtained originally by 
 \cite{Gunion:1973ex}   by using time-ordered perturbation theory in
 the infinite-momentum-frame IMF. Thus three different approaches
 yield the same exact result for this model problem. The asymptotic limit of
 asymptotically high momentum transfer $Q^2$ is also studied with the result
 that  $F(Q^2)\sim 1/2\ln^2 Q^2/Q^2$  The next section (IV) explains 
 the meaning of transverse density $\rho(b)$ of the model. 
Its central value varies singularly  as $\ln^3(b)/3$. A general
 derivation of the relationship of $\rho(b)$  with the form factor using three
 dimensional spatial coordinates is presented. This allows us to
 identify a mean-square transverse size $\langle b^2\rangle=\int
 d^2b\;b^2\rho(b)$ 
that is given by $b^2=-4{dF\over dQ^2}(Q^2=0)$. The quantity  $\langle
b^2\rangle$ is a true measure of hadronic size because of its direct
relationship with the transverse density. Using this model it is
possible to 
  display the spatial wave function in terms of three
spatial coordinates (Section~V), but this is not very useful.
 Section~VI shows that the rest-frame charge
distribution is generally not observable by studying the 
explicit failure to uphold current conservation.  Section~VII shows
that  neutral systems of two constituents
obey the conventional lore that the heavier one is generally closer to
the transverse origin than the lighter one. It is also argued that 
the negative central charge density of the neutron arises in
pion-cloud models from pions residing at the center of the nucleon.  
 The non-relativistic limit is defined
and applied to a variety of examples in
Section~VIII. By varying the masses one can study a continuum of
examples in which the constituents move at a wide range of average
velocities. The relevant quantity is the ratio of the binding energy
$B$ to that of the mass ${\cal M} $ of the lightest constituent ($\phi$ or
$\xi$). For small values of $B/{\cal M}$ 
the exact relativistic formula is shown to be the same as 
the
familiar one of the three-dimensional Fourier transform of a square of
a wave function.  If the $\phi$ and $\xi$ have equal masses $m$ we
find that $B/(2m)$ must be less than 0.001 for the relativistic
corrections to mean-square radii to be be less than 10\%, see
\eq{eqm}.
For the case when $m_\xi=2m_\phi$ which mimics the quark-di-quark
model of the nucleon we find that there are substantial relativistic
corrections to the form factor for any value of $Q^2$. 
 This means that one can not take a
three-dimensional Fourier transform of the form factor to get a charge
density even if the constituent masses are large. A schematic
model of the lowest $s$-states of nuclei is developed by choosing
$m_\xi=(A-1)m_\phi$, where $A$ is the nucleon number. Relativistic
effects are found to decrease the form factor for light nuclei but to 
increase the form factor for heavy nuclei. Furthermore, these
lowest $s$-states are likely to be strongly influenced by relativistic
effects that
are order 15-20\%.  

I thank the USDOE (FG02-97ER41014)
for partial support of this work, S. Brodsky for advocating the
use of the $\phi^3$ model as a pedagogic tool, and 
 J. Arrington, A. Bernstein,  M. Burkardt, I. Clo\"et, B. Jennings, E.
 Henley, and W. Polyzou
for useful discussions.
 

\begin{thebibliography}{0}
\bibitem{Hofstadter:1956qs}
  R.~Hofstadter,
  Rev.\ Mod.\ Phys.\  {\bf 28}, 214 (1956).

\bibitem{Weinberg:1966jm}
  S.~Weinberg,
  Phys.\ Rev.\  {\bf 150}, 1313 (1966).


\bibitem{Gunion:1973ex}
  J.~F.~Gunion, S.~J.~Brodsky and R.~Blankenbecler,
  Phys.\ Rev.\  D {\bf 8}, 287 (1973).

\bibitem{reviews} 
 H.~y.~Gao,
  Int.\ J.\ Mod.\ Phys.\  E {\bf 12}, 1 (2003)
  [Erratum-ibid.\  E {\bf 12}, 567 (2003)];
 C.~E.~Hyde-Wright and K.~de Jager,
  Ann.\ Rev.\ Nucl.\ Part.\ Sci.\  {\bf 54}, 217 (2004);
  C.~F.~Perdrisat, V.~Punjabi and M.~Vanderhaeghen,
Prog.\ Part.\ Nucl.\ Phys.\  {\bf 59}, 694 (2007);
 J.~Arrington, C.~D.~Roberts and J.~M.~Zanotti,
  J.\ Phys.\ G {\bf 34}, S23 (2007).

\bibitem{Miller:2007uy}
  G.~A.~Miller,
  Phys.\ Rev.\ Lett.\  {\bf 99}, 112001 (2007).

\bibitem{notation}
  Our notation, except for Sect.~III,
 is that $x^\pm\equiv (x^0\pm x^3)/\sqrt{2},p^\pm\equiv (p^0\pm p^3)/\sqrt{2}$, and
$p_\mu x^\mu=p^-x^++p^+x^--\bfp\cdot\bfb$. In Sect.~III we use $p^\pm =
p^0\pm p^3$.
The coordinates perpendicular to the 0 and 3 directions are denoted as
$\bfb$ and $\bfp$.

\bibitem{soper1} D.E. Soper, Phys.\ Rev.\ D\ {\bf 5},
 1956 (1972). 

\bibitem{mbimpact}
  M.~Burkardt,
  Int.\ J.\ Mod.\ Phys.\  A {\bf 18}, 173 (2003).


\bibitem{diehl2} 
  M.~Diehl,
  Eur.\ Phys.\ J.\  C {\bf 25}, 223 (2002)
  [Erratum-ibid.\  C {\bf 31}, 277 (2003)].
\bibitem{Carlson:2008zc}
  C.~E.~Carlson and M.~Vanderhaeghen,
  Phys.\ Rev.\ Lett.\  {\bf 100}, 032004 (2008)



\bibitem{miller:2009qu}
G.~A.~Miller, 
Phys.\ Rev.\  C {\bf 79}, 055204 (2009) 


\bibitem{mb1} M. Burkardt, Phys.\ Rev.\ D\
{\bf 62}, 071503 (R) (2000).


\bibitem{diehl} M. Diehl {\it et al.}, Nucl.\ Phys.\
B\ {\bf 596}, {33} (2001).


\bibitem{Miller:2007kt}
  G.~A.~Miller, E.~Piasetzky and G.~Ron,
  Phys.\ Rev.\ Lett.\  {\bf 101}, 082002 (2008)






\bibitem{Pirner:2009zz}
  H.~J.~Pirner, B.~Galow and O.~Schlaudt,
  Nucl.\ Phys.\  A {\bf 819}, 135 (2009).
 H.~J.~Pirner and N.~Nurpeissov,
  Phys.\ Lett.\  B {\bf 595}, 379 (2004.)



 
\bibitem{Miller:2008jc}
  G.~A.~Miller and J.~Arrington,
  Phys.\ Rev.\  C {\bf 78}, 032201 (2008).
\bibitem{Rinehimer:2009yv}
  J.~A.~Rinehimer and G.~A.~Miller,
 Phys. Rev. C 80, 015201 (2009)
\bibitem{Rinehimer:2009sz}
  J.~A.~Rinehimer and G.~A.~Miller,
  arXiv:0906.5020 [nucl-th].
\bibitem{assum}The use of
 $m_1=M$ involves neglecting the effect of the pion cloud in shifting the nucleon
mass.
\bibitem{Strikman:2009bd}
  M.~Strikman and C.~Weiss,
  arXiv:0906.3267 [hep-ph].



%
\bibitem{Oettel:2000jj}
   M.~Oettel, R.~Alkofer and L.~von Smekal,
   Eur.\ Phys.\ J.\  A {\bf 8}, 553 (2000).

\bibitem{Horikawa:2005dh}
   T.~Horikawa and W.~Bentz,
   Nucl.\ Phys.\  A {\bf 762}, 102 (2005).

\bibitem{Cloet:2008re}
   I.~C.~Clo\"et, G.~Eichmann, B.~El-Bennich, T.~Klahn and C.~D.~Roberts,
   Few Body Syst.\  {\bf 46}, 1 (2009).


\bibitem{Brodsky:1989pv}
  S.~J.~Brodsky and G.~P.~Lepage,
  Adv.\ Ser.\ Direct.\ High Energy Phys.\  {\bf 5}, 93 (1989).

\bibitem{Frankfurt:1981mk}
  L.~L.~Frankfurt and M.~I.~Strikman,
  Phys.\ Rept.\  {\bf 76}, 215 (1981).


\end{thebibliography}
\end{document}